\documentclass[journal]{IEEEtran}
\ifCLASSINFOpdf
\else
\fi

\usepackage{ulem} 
\usepackage{amsmath}
\usepackage{graphicx}
\usepackage{subfigure}
\usepackage{color}
\usepackage{amsthm,amsmath,amssymb}
\usepackage{mathrsfs}
\usepackage{ulem}
\usepackage{amsmath} 
\usepackage{amssymb}
\usepackage{algorithm}
\usepackage{algorithmic}
\usepackage{color}
\usepackage{verbatim}
\usepackage{float}
\usepackage{graphicx}
\usepackage{url}
\usepackage{algorithm}
\usepackage{algorithmic}
\usepackage{color}
\usepackage{verbatim}
\usepackage{float}
\usepackage{cite}
\usepackage{caption}

\begin{document}
\title{Task-Load-Aware Game-Theoretic Framework for Wireless Federated Learning}

\author{Jiawei Liu, Guopeng Zhang, Kezhi Wang, and Kun Yang
\vspace{-3mm}
\thanks{Jiawei Liu and Guopeng Zhang are with the School of Computer Science and Technology, China University of Mining and Technology, Xuzhou 221116, China (e-mail: jwliu@cumt.edu.cn; gpzhang@cumt.edu.cn).}

\thanks{Kezhi Wang is with the Department of Computer and Information Science, Northumbria University, Newcastle NE1 8ST, U.K. (e-mail: kezhi.wang@northumbria.ac.uk).}

\thanks{Kun Yang is with the School of Computer Science and Electronic
Engineering, University of Essex, Colchester CO4 3SQ, U.K. (e-mail:
kunyang@essex.ac.uk).} }

\maketitle

\begin{abstract}
Federated learning (FL) can protect data privacy but has difficulties in motivating user equipment (UE) to engage in task training. This paper proposes a Bertrand-game based framework to address the incentive problem, where a model owner (MO) issues an FL task and the employed UEs help train the model by using their local data. Specially, we consider the impact of time-varying \textit{task load} and \textit{channel quality} on UE's motivation to engage in the FL task. We adopt the finite-state discrete-time Markov
chain (FSDT-MC) to predict these parameters during the FL task. Depending on the performance metrics set by the MO and the estimated energy cost of the FL task, each UE seeks to maximize its profit. We obtain the Nash equilibrium (NE) of the game in closed form, and develop a distributed iterative algorithm to find it. 
Finally, simulation result verifies the effectiveness of the proposed approach.
\end{abstract}


\section{Introduction}
\IEEEPARstart{F}{ederated} learning (FL) \cite{2016Federated} is a distributed machine learning (ML) framework that allows multiple clients to collaboratively train an ML model without exposing their raw data. In FL framework, a model owner (MO) first sends a global model to edge or mobile clients. Then the clients can train the model with their local data and transmit the updated model parameters to the MO for a new aggregated model. This process iterates until the required accuracy is achieved.

Although FL protects data privacy, it causes additional energy cost of clients for model training and parameter transmission, thus limiting their motivation to engage in FL \cite{9409833}. To address this issue,  a Stackelberg game is proposed in \cite{8867906} to encourage clients to reduce the completion time of FL tasks. In \cite{8963610}, a reinforcement learning-based incentive mechanism is designed to find the optimal pricing and training strategies for the clients. In \cite{8995775}, a Stackelberg game is proposed to motivate the MO and clients to build a high-quality FL model. However, several strong assumptions were made in the above works, e.g., the clients have invariant \textit{task loads} and \textit{channel quality} during an FL task.
These assumptions apply only to situations where the clients are computationally powerful edge servers connected with the wired networks. 
When the clients are mobile UEs, the energy and computation power are normally limited, thus the influence of time-varying \textit{task load} and \textit{channel quality} on their motivation to engage in FL may be considered.

Against the above background, this paper proposes a Bertrand-game\cite{1988game} based framework to motivate UEs to engage in FL \cite{decentralized} \cite{dynamicgame}, which considers the changes of \textit{task load} and \textit{channel quality} of UEs during an FL task. Firstly, given the performance metrics set by the MO, we show that the energy cost of a UE depends not only on the \textit{FL-task load} but also on its \textit{existing-task load}. We characterize the dynamic of user's \textit{future task load} as an FSDT-MC \cite{fsmccom}. Then, the energy cost of a UE caused by multiple rounds of local training can be predicted. Furthermore, we employ another FSDT-MC to describe the \textit{channel fading} between a UE and the MO, as it has been widely used to simulate Rayleigh fading channels \cite{fsmccommunication}. Then, the energy cost of the UE caused by multiple rounds of parameter transmission can also be estimated. Depending on the predicted overall energy cost for model training and parameter transmission, the UEs can charge the MO independently for the resource usage. At the NE of the game, the MO achieves a set of specified performance metrics with the minimum total payment, while each UE seeks the best price to maximize its own profit. The contribution of this paper is as follows:

1) By integrating the FSDT-MC based \textit{task load} and \textit{channel quality} prediction method, a Bertrand game is proposed to help motivate UEs to engage in FL based on the estimated overall energy cost.

2) The NE of the game is addressed in closed form, and an effective iterative algorithm is designed to obtain the NE in a distributed manner. 

\section{System Model}
The considered FL system consists of one MO and a set $\mathcal{K}$ of $K$ mobile UEs. The MO can communicate with the UEs via direct wireless links. Each UE $k$ ($\forall k\in\mathcal{K}$) stores a local dataset $\mathcal{D}_k$ of size $|\mathcal{D}_k|$, and the $i^\text{th}$ data sample in set $\mathcal{D}_k$ is represented by $d_{k,i} = \left \{\textbf{x}_{k,i},y_{k,i}\right\}$, $\forall d_{k,i} \in \mathcal{D}_k$. Let $\phi\left(d_{k,i};\textbf{w}\right)$ denote the global loss function of the MO, where $\textbf{w}$ is the model parameter. To minimize $\phi\left(d_{k,i};\textbf{w}\right)$ without sharing the data of the UEs, the MO can use the FL algorithm \texttt{FedAVG} \cite{2016Federated} as given below.

\textit{Step 1)} The MO broadcasts the initial model parameter $\textbf{w}$ to all the UEs in set $\mathcal{K}$.

\textit{Step 2)} Each UE $k$ can continue to train $\textbf{w}$ on the local dataset $\mathcal{D}_k$ and solve the local optimal model $\textbf{w}_k$ by addressing the following problem \cite{pingfangxiang}
\begin{equation}
        \textstyle \min_{\textbf{w}_k} \frac{1}{|\mathcal{D}_k|} \sum_{i=1}^{ |\mathcal{D}_k|} \phi_k\left(d_{k,i};\textbf{w}_k\right), \ \forall d_{k,i} \in \mathcal{D}_k. \label{lossfunction1}
\end{equation}

On condition that $\phi_k(d_{k,i};\textbf{w}_k)$ is convex, problem \eqref{lossfunction1} can be solved by using an iterative approach. Let $\textbf{w}_k^*$ denote the optimal solution to problem \eqref{lossfunction1}. From \cite{2016Federated}, we know that UE $k$ can use arbitrary optimization algorithms (such as the Stochastic Gradient Descent (SGD)) to attain a relative accuracy $\theta_k$ for problem \eqref{lossfunction1}, such that $\mathbb{E} \left[\phi_k\left(d_{k,i};\textbf{w}_k\right)-\phi_k(d_{k,i};\textbf{w}_k^*)\right] \leq \theta_k$. To achieve $\theta_k$, the general upper bound on the required number of local iterations $I_k$ is given by \cite{2016Federated}
\begin{equation}
    I_k(\theta_k)=\eta_k \log \left(1/\theta_k\right), \ \forall k\in\mathcal{K}, \label{localiter}
\end{equation}
where $\eta_k \geq 0$ is a parameter set by UE $k$.

\textit{Step 3)} After each round of local training, each UE $k$ transmits the updated $\textbf{w}_k$ to the MO via wireless link. The MO can update the global model as $\textbf{w} = \sum_{k=1}^{K}\frac{|\mathcal{D}_k|}{|\mathcal{D}|}\textbf{w}_k$, where $\mathcal{D}$ = $\bigcup_{k=1}^{K}\mathcal{D}_k$, and then broadcasts the updated $\textbf{w}$ to all the UEs to repeat the above process.

After several rounds of the global update, the MO can find the model parameter $\textbf{w}$ to minimize the global loss function $\phi\left(d_{k,i};\textbf{w}\right)$.
Let $\textbf{w}^*$ denote the optimal solution to the global problem. We use $\epsilon$ ($0\leq \epsilon \leq 1$) to represent the global relative accuracy which characterizes the quality of the solution $\textbf{w}$ to the global problem as it produces a random output satisfying $\mathbb{E}[\phi \left(d_{k,i};\textbf{w} \right)-\phi\left(d_{k,i};\textbf{w}^*\right)] \leq \epsilon$. 
Due to the heterogeneity of the UEs, the upper bound of the number of global iterations $I^\text{g}$ to implement $\epsilon$ is given by \cite{2016Federated}
\begin{align}
    I^{\text{g}}(\epsilon,\theta_k)=\frac{\zeta\log (\frac{1}{\epsilon})}{1-\max_{k}\theta_k}, \ \forall k\in\mathcal{K},\label{I^g}
\end{align}
where $\zeta > 0$ is a constant specified by the MO. 


\subsection{Energy cost for model training}
\begin{figure}[ht]
  \centering
  \includegraphics[scale=0.4]{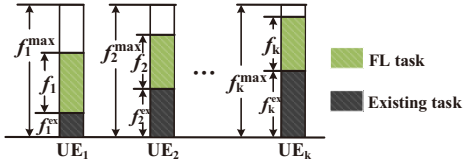}
  \vspace{-2mm}
  \caption{The computational resource allocation of the UEs.} 
  \label{systemfig}
\end{figure}
Before engaging in an FL task, each UE $k$ may process some \textit{existing tasks} (termed as EX-tasks in the following analysis) as shown in Fig. \ref{systemfig}.
We use $f_{k,t}^\text{ex}$ to represent the CPU frequency of UE $k$ to process the EX-task, and assume that $f_{k,t}^\text{ex}$ remains constant during the local training session $t$ ($1\leq t \leq I^\text{g}$) whose duration is $T^{\text{trn}}$. To admit the FL-task, UE $k$ requires an additional $C_k=c_k |\mathcal{D}_k| I_k$ CPU cycles during each session $t$, where $c_k$ is the number of CPU cycles required for training per data sample $d_{k,i}$. Therefore, UE $k$ should raise the CPU frequency by $
f_k= C_k/ T^{\text{trn}}$ Hz in each session to complete the newly carried FL-task. According to \cite{pingfangxiang}, the additional energy cost of UE $k$ during session $t$ is given as
\begin{equation}
    E_{k,t}^{\text{F}}=\nu_k \left(f_{k,t}^\text{ex} + f_k\right)^2T^{\text{trn}}- \nu_k \left(f_{k,t}^\text{ex}\right)^2 T^{\text{trn}},  \label{energytrn_1} \end{equation}
where $\nu_k$ is the effective switched capacitance of the CPU. 
Let $f^{\text{max}}$ denote the maximum CPU frequency of the UEs. The following constraint should be satisfied.
\begin{equation}
   0 < f_{k,t}^\text{ex}+f_k \leq f^\text{max},\ \forall k\in{\mathcal{K}},\ \forall t \in \mathcal{T}.
\end{equation}

From eq. \eqref{energytrn_1}, we know that $E_{k,t}^{\text{F}}$ depends not only on the FL-task load $f_k$ but also on the EX-task load $f_k^\text{ex}$, which causes the UEs to produce the same amount of CPU cycles, i.e., $C_k$ but consume different amount of energy $E_{k,t}^{\text{F}}$. It will affect the pricing strategy of the UEs in the game. It is noted that the fluctuation of the task load in the UE over a short period (e.g., a few tens of seconds) has some kind of randomness, but the task loads in the adjacent time periods may have a strong correlation \cite{2021resource}. Next, we propose an FSDT-MC based method \cite{fsmccom} to predict the dynamic of $f_{k,t}^\text{ex}$ during the FL period. Therefore, each UE can estimate its energy cost for model training and set the price for the resource usage.

    \begin{enumerate}
        \item From ref. \cite{2021resource}, we know that the workload fluctuation of a machine over a longer period (e.g., a few minutes) has a strong periodicity, while the workload fluctuation over a shorter period (e.g., a few tens of seconds) has a certain randomness, but the workloads in the adjacent time periods have a strong correlation.
        \item In this paper, the time spent for a single round of local training is $T^\text{trn}=2$s, the time spent for a single round of parameter transmission is $T^\text{com}=0.2$s, and the total number of local training sessions is $I^\text{g}=10$. Therefore, the total time required to complete a federated learning (FL) task is $T=I^\text{g}(T^\text{trn}+T^\text{com})=22$s.
        \item In the proposed game, the UEs must determine the price of the consumed resource during a FL task.
        In order to capture the short-term dynamic of users' workload during the task, we set the \textit{observation interval} and \textit{update period} of $Q_k^\text{F}$ to $T=22$s (as stated in the response to your last question), and $Q_k^\text{F}$ can be timely updated by the workload changes collected at runtime.        
    \end{enumerate}
    
According to \cite{fsmccom}, we discretize the value space of $f_{k,t}^\text{ex}$ ($0\leq f_{k,t}^\text{ex}\leq f^\text{max}$) into $M$ equal-width bins as $\textbf{F}=\{F_1,\ \cdots,\ F_M\}$, where $F_m=\frac{m-1}{M-1}f^\text{max}$ ($1\leq m\leq M$) represents a state of the FSDT-MC. Any UE $k$ can learn and update the state transition probability (STP) matrix $Q_k^\text{F}=\left(\alpha_{m, m'}\right)_{M\times M}$ every $T$ seconds by using the statistical method , where $Q_k^\text{F}$ is an $M\times M$ matrix and $\alpha_{m, m'}$ is the conditional probability of transitioning from state $m$ to $m'$. To reflect the short-term dynamics of $f_{k,t}^\text{ex}$ during an FL period, we set the update period of $Q_k^\text{F}$ to $T=I^\text{g}(T^\text{trn}+T^\text{com})$. The method to obtain and update $Q_k^\text{F}$ as follows.

  \begin{enumerate}
        \item Define the \textit{observation interval} or \textit{update period} of $Q_k^\text{F}$ as $T=I^\text{g}(T^\text{trn}+T^\text{com})$, where $I^\text{g}$ represents the number of local training sessions, and $T^\text{trn}$ ($T^\text{com}$) represent the time spent for a single round of local training (parameter transmission). It means that $T$ the \textit{observation interval} or \textit{update period} of $Q_k^\text{F}$  is the total time required to complete a federated learning task.
        \item We divide the \textit{observation interval} $T$ into a large number of (e.g., $K$) small \textit{time slots} with equal length $\tau=T/K$.
        We can use the operating system of a UE to monitor and record the workload, i.e. the operating frequency of the CPU in each slot.
        \item Let $N_{m}^{T}$ denote the number of times that the CPU frequency is in state $m$ during the \textit{observation interval}, $N_{m,i}^T$ the number of times that the CPU frequency changes from state $m$ to state $i$ during the \textit{observation interval}. According to the following literature \cite{2013tongji}, we can use the observed records to obtain the frequency of the change from state $m$ to state $i$ during the \textit{observation interval} as
        \begin{equation*}
            \hat{p}_{m,i}=\frac{N_{m,i}^T}{N_{m}^{T}},\ 1\leq i,m\leq M,\ \text{and}\ m \neq i, 
        \end{equation*}

        Then, we can take $\hat{p}_{m,i}$ as the probability of the change from state $m$ to state $i$ during the \textit{observation interval}, that is 
\begin{equation*}
 \alpha_{m,i}=\hat{p}_{m,i}=\frac{N_{m,i}^T}{N_{m}^T},\ 1\leq i,m\leq M,\ \text{and}\ m \neq i.
            \label{probability}
\end{equation*}

\item According to the principle of Markov chain, i.e. the next state of transition only depends on the current state and is independent on the previous states, the transition matrix $Q_k^\text{F}$ of UE $k$ during local training session $t$ is constructed as
        \begin{equation*}
            Q_k^\text{F} = \left( \alpha_{m,i}\right)_{M\times M},\ 1\leq i,m\leq M,\ \text{and}\ m \neq i.
        \end{equation*}
    \end{enumerate}
    
Suppose that $f_{k,t-1}^\text{ex}$ is in state $m$ during session $(t-1)$. UE $k$ can predict that $f_{k,t}^\text{ex}$ is in state $m_t$ during session $t$ by extracting the most probable future state $m'$ (i.e., the one with the largest STP) from $Q^\text{F}_k$ at row $m$, which is
\begin{equation}
 \textstyle m_t=\mathop{\arg\max}\limits_{\forall m'\in\{1,\cdots,M\}} \alpha_{m,m'},\ \forall t \in{\mathcal{T}}. \label{etl_state}
\end{equation}

By substituting eq. \eqref{etl_state} into eq. \eqref{energytrn_1}, UE $k$ can estimate its energy cost in training session $t$ as
\begin{equation}
    \tilde{E}_{k,t}^{\text{F}}=\nu_k \left(F_{m_t} + f_k\right)^2T^{\text{trn}}- \nu_k \left(F_{m_t}\right)^2 T^{\text{trn}}, \ \forall t \in{\mathcal{T}}. \label{energytrn_2}
\end{equation}

\subsection{Energy cost for parameter transmission}
After each round of local training, the MO assigns UE $k$ a time period $T^\text{com}$ and an orthogonal channel of $W$ Hz to upload the updated model parameter $\textbf{w}_k$. Let $L$ denote the size of $\textbf{w}_k$, $p_{k,t}$ the transmit power of UE $k$, and $g_{k,t}$ the channel gain from UE $k$ to the MO during the $t^{\text{th}}$ parameter transmission. Then, the achievable data rate of UE $k$ under a given bit error rate (BER) level can be approximated as
\begin{equation}
    r_{k,t}=W\log_2\left(1+ {\Delta  p_{k,t} g_{k,t}}/\sigma^2\right),
\end{equation}
where $\sigma^2$ is noise power and $\Delta =\frac{1.5}{-\ln(5\text{BER})}$ is a constant related to BER. Since $r_{k,t} T^\text{com}\geq L$ should be satisfied, the minimum transmit power of UE $k$ is derived as 
\begin{equation}
p_{k,t} = \left(2^{\frac{L}{WT^\text{com}} }-1\right) {\sigma^2}/(g_{k,t}\Delta ).
\end{equation}
and the resultant energy cost of UE $k$ is
\begin{align}
   E_{k,t}^{\text{C}}= p_{k,t} T^\text{com}= \left(2^{\frac{L}{WT^\text{com}} }-1\right){\sigma^2}  T^\text{com}/(g_{k,t}\Delta ). \label{energycom_01}  
\end{align}

Next, we construct an FSDT-MC \cite{fsmccommunication} to predict $g_{k,t}$ ($\forall t \in{\mathcal{T}}$) for the UEs to develop pricing strategies.

Let $T_c$ denote the channel correlation time, over which the channel response is invariant. For instance, when the system operates on the 900 MHz of 4G (wavelength $\lambda=0.34$ m) and the UEs move slowly (speed $v\approx 2 $ m/s), $T_c=\frac{\lambda}{v}\approx0.2$s \cite{fsmccommunication}.
We set $T^\text{com} = T_c $ and $T^\text{trn} = \delta T_c$, where $\delta \gg 1$ is a positive integer.
Divide the time horizon of the system into time slots with equal length $T_c$ and index the slots by $\tau$. Then, $g_{k,\tau}$ is exponentially distributed over $\tau$. We represent the lower and upper bounds of $g_{k,\tau}$ by $\check{g}$ and $\hat{g}$, respectively, and can then discretize the value space of $g_{k,\tau}$ into $N$ equal-width levels as $\textbf{L}=\{L_1,\ \cdots,\ L_N\}$, where $L_n=\frac{n-1}{N-1}(\hat{g}-\check{g})$, $1\leq n\leq N$, represents a state of the FSDT-MC \cite{fsmccommunication}.

Define the STP matrix of $g_{k,\tau}$ over $\tau$ as $Q^\text{C}=\left(\beta_{n, n'}\right)_{N\times N}$, where $\beta_{n, n'}$ is the probability that $g_{k,\tau}$ changes from state $n$ to $n'$ as the time goes from $\tau$ to $\tau+1$. 
We can obtain $Q^\text{C}$ by using the statistical method. The main idea is to take the frequency of $g_{k,\tau}$ changing from state $n$ to $n'$ during an observation/update period as the STP $\beta_{n, n'}$. The update period can be adjusted adaptively to reflect the change rate of the channel.

 \begin{enumerate}
        \item Let $\check{g}$ and $\hat{g}$ denote the lower and upper bounds of $g_{k,\tau}$, respectively. We discretize the value of $g_{k,\tau}$ into $N$ equal-width levels as $\textbf{L}=\{L_1,\ \cdots,\ L_N\}$, where
        \begin{equation*}
            L_n=\frac{n-1}{N-1}(\hat{g}-\check{g}), 1\leq n\leq N    
        \end{equation*}
        represents a state of the FSDT-MC. The state transition of $g_{k,\tau}$ over $\tau$ can be described by matrix $Q^\text{C}=\left(\beta_{n, j}\right)_{N\times N}$, where $\beta_{n, j}$ is the probability that $g_{k,\tau}$ changes from state $n$ to state $j$ as the time goes from $\tau$ to $\tau+1$.

        \item Let $O=X\tau$ denote the \textit{observation period} or \textit{update period} of $Q^\text{C}$, where $X$ is a sufficiently large positive integer. We use $Z_n$ and $Z_{n,j}$ to represent the numbers of times per {observation period} that $g_{k,\tau}$ stays in level $L_n$ and changes from level $L_n$ to level $L_j$, respectively. Following the priciple of Markov chain, the transition probabilities from state $L_n$ to state $L_j$, i.e., $\beta_{n,j}$, can be approximated by the ratio
        \begin{equation*}
            \beta_{n,j}=\frac{Z_{n,j}}{Z_n}, \ 1\leq n,j\leq N,\  n\neq j.
        \end{equation*}
        As a result, the transition matrix $Q^\text{C}=\left(\beta_{n, j}\right)_{N\times N}$ is obtained.

        \item When a user moves not so fast, i.e, the channel state is relatively stable, we can increase the update period of $Q^\text{C}$ by increasing $X$. However, when the user moves faster and the channel state changes more dramatically, we can reduce the update period of $Q^\text{C}$ by reducing $X$, thus reflecting the rapid change of the channel state. In this paper, we set $X=100$ considering that walking users move at a low speed of 2 m/s. 
\end{enumerate}

Assuming that the initial distribution of $g_{k,\tau=0}$ over the $N$ states is known as $\pi(g_{k,\tau=0})_{1\times N}$, UE $k$ can predict that $g_{k,t}$ is in state $n_t$ during the $t^\text{th}$ parameter transmission by extracting the most probable future state $n'$ from $Q^\text{C}$ at row $n$, which is
\begin{equation}
    n_t=\mathop{\arg\max}\limits_{\forall n'\in\{1,\cdots,N\}} \pi(g_{k,\tau=0})(Q^\text{C})^{t(\frac{T^\text{trn}}{T_c}+1)},\ \forall t \in{\mathcal{T}}. \label{chgain}
\end{equation}
By substituting eq. \eqref{chgain} into eq. \eqref{energycom_01}, UE $k$ can estimate its energy cost for the $t^\text{th}$ parameter transmission as
\begin{align}
   \tilde{E}_{k,t}^{\text{C}}= (2^{\frac{L}{WT^\text{com}} }-1){\sigma^2}  T^\text{com}/(L_{n_t}\Delta ),\ \forall t \in{\mathcal{T}}. \label{energycom_02}
\end{align}
where $L_{n_t}$ is the $n_t^\text{th}$ state of the FSDT-MC of the channel.

\section{Game Formulation}
To motivate the UEs to engage in FL, an effective way is to enable the UEs to profit from the energy consumption of model training and data uploading. Game theory is a powerful tool to analyze the optimal or equilibrium strategies of the UEs in such a competitive situation. In general, strategic games can be divided into the following two categories \cite{1988game}: (1) the production-based games, such as \textit{Cournot} and \textit{Stackelberg games}, which take the \textit{supply quantity} of a commodity as a strategy to maximize the players' profits, and (2) the price-based games, such as \textit{Bertrand games}, which take the \textit{price} of a commodity as a strategy to maximize the players' profits. In the FL system, the UEs have no access to the amount of resources that other UEs contribute, but they can observe the pricing information of others to make decisions. Thus we adopt the Bertrand game to solve the incentive problem. While the UEs fight for market share through price competition, the MO can adjust its purchase according to the prices offered by the UEs, so as to achieve its performance with a minimum total investment. Next, we model the price competition between the UEs and their interaction with the MO as a Bertrand game. The \textit{game execution cycle} is the time to complete an entire FL task with a continuous $I^{\text{g}}$-round global iterations.

\begin{enumerate}
    \item Strategic games can be divided into the following \textit{quantity}-based games and \textit{price}-based games [6].

    \begin{enumerate}
        \item Games based on \textit{commodity production}, such as \textit{Cournot games} and \textit{Stackelberg games}, in which competing monopolists take the supply quantity of a commodity as a strategy to maximize their profits.
        \item Games based on \textit{commodity price}, such as \textit{Bertrand games}, in which competing monopolists take the price of a commodity as a strategy to maximize their profits.
    \end{enumerate} 
    In general, the monopolists cannot obtain each other's production information, but they can easily observe the rivals' product pricing from the market, which helps the monopolists to make decisions. The price of a commodity is a signal to coordinate the supply and demand of the commodity in the market. 

    \item In this paper, we motivate the UEs to participate in FL through pricing the consumed resources, and, therefore, the \textit{Bertrand game} is adopted to solve problem (14).
    \begin{enumerate}      
        \item In problem (14), multiple competitive UEs produce homogeneous products (computing and communication resources) and set the prices based on their specific production costs (determined by their respective energy consumption during FL) and others' prices to meet the market demand (the FL performance ordered by the model server).
        \item Similar to the monopolists in a oligopoly market, the UEs have no access to the amount of resources that others contribute to the FL task, but they can obtain the price-information of others to make decisions.
        Therefore, the proposed game belongs to the price-based games and we adopt the \textit{Bertrand game} to analyze the incentive problem (14).
    \end{enumerate}
\end{enumerate}

\textit{1) The objective of UEs:} We represent the price asked by UE $k$ for training session $t$ by $\varrho_{k,t}$ (in \textit{Joule/iteration}), which is the compensation required by UE $k$ for each local iteration over the dataset $\mathcal{D}_k$.
Denote the overall energy cost of UE $k$ for completing a round of local training and parameter transmission by $\psi_{k,t} =\tilde{E}_{k,t}^\text{F}+\tilde{E}_{k,t}^\text{C}$, and the payment of the MO to UE $k$ by $\xi_{k,t}= \varrho_{k,t} I_k$. The aim of UE $k$ is to maximize its profit in the total $I^{\text{g}}$ rounds by choosing the optimal price profile $\Gamma_k=\left(\varrho_{k,1},\cdots,\varrho_{k,I^\text{g}}\right)$. This can be formulated as the following problem.
\begin{alignat}{3}
    &\max_{\Gamma_k} \; && U_k=\sum_{t\in \mathcal{T}}(\xi_{k,t}-\psi_{k,t})=\sum_{t\in \mathcal{T}}(\varrho_{k,t} I_k - (\tilde{E}_{k,t}^\text{F}+\tilde{E}_{k,t}^\text{C})) & \label{maxUK} \\
    &\mbox{s.t.} && \varrho_{k,t} \geq 0, \; \forall k \in{\mathcal{K}}, \forall t \in{\mathcal{T}}. & \tag{\ref{maxUK}.1} \label{positiveq}
\end{alignat}
\textit{2) The objective of the MO:} The MO aims to achieve its performance (including the global accuracy $\epsilon$ and the training parameters $I^{\text{g}}$, $T^{\text{trn}}$ and $T^\text{com}$) with a minimum total investment. Let $\Theta=\left(\theta_1,\cdots,\theta_{\left|\mathcal{K}\right|}\right)$ denote the strategy profile of the MO, which represents the amount of resources it purchases from each UE $k$. Once the UEs determine their optimal prices $\Gamma_k,\  \forall k \in \mathcal{K}$, the MO can adjust $\Theta$ to achieve its objective, which is formulated as the following problem
\begin{alignat}{3}
&\min_{\Theta} \; && U_{\text{S}}= \sum_{t\in \mathcal{T}} \sum_{k\in K} \varrho_{k,t} I_k +\frac{1}{2}\Big(\sum_{k\in K}(\theta_k)^2 +2v\sum_{k \neq j}(\theta_k \theta_j )\Big), \label{minU} \\
    &\mbox{s.t.} && 0 < \theta_k \leq \theta^\text{max} =1-\frac{\zeta\log (\frac{1}{\epsilon})}{I^{\text{g}}}, \, \; \forall k\in{\mathcal{K}}, & \tag{\ref{minU}.1} \label{positivel}
\end{alignat}
where constraint \eqref{positivel} indicates the value space of $\theta_k$. From eq. \eqref{I^g}, we can get $\theta^\text{max}=\max_{k}\theta_k=1-\frac{\zeta\log (\frac{1}{\epsilon})}{I^{\text{g}}}$. The meaning of $U_{\text{S}}$ is given as follows.

\begin{enumerate}
    \item \textit{The meaning of the first term $\sum_{t\in \mathcal{T}} \sum_{k\in K} \varrho_{k,t} I_k$}.
    \begin{enumerate}
        \item Parameter $I_k$ represents the required number of local iterations in each $k^\text{th}$ training session to achieve local accuracy $\theta_k$. 
        
        \item Parameter $\varrho_{k,t}$ represents the price asked by UE $k$ for the $t^\text{th}$ training session.
        
        \item Therefore, $\sum_{t\in \mathcal{T}} \sum_{k\in K} \varrho_{k,t} I_k$ represents the total payment of the server to all the UEs in the total $I^{\text{g}}$ rounds of local training.
    
    \end{enumerate}
    
    \item The second term $\frac{1}{2}\Big(\sum_{k\in K}(\theta_k)^2 +2v\sum_{k \neq j}(\theta_k \theta_j )\Big)$ considers the \textit{Resource Substitutability} (RS) in a Bertrand game.
    
    \begin{enumerate}
    \item We note that the Bertrand game model is first proposed in \cite{tidaicanshu}. The content related to the RS is in the second formula, Part 2, page 547 of ref. \cite{tidaicanshu}. Next, we introduce it briefly.
    
    \item According to ref. \cite{substitutability}, the RS is an ability of a buyer to substitute one product (hold by some sellers) with other products (hold by other sellers) of similar functionality. The RS of a buyer is described by parameter $v \in \left[0, 1\right]$. When $v = 0$, there is no substitutability between products sold by different sellers, when $v = 1$, the products sold by different sellers are completely homogeneous with each other. Please refer to it for more details.
    
    \item In what follows, we explain how the RS is applied to our game framework for federated learning.
    
    The function of term $\frac{1}{2}\Big(\sum_{k\in K}(\theta_k)^2 +2v\sum_{k \neq j}(\theta_k \theta_j )\Big)$ can be explained from the economic aspect, as it tasks the RS into account through parameter $v$. 
    
    It is known that a major challenge to implement FL is to solve the non independent-and-identical distribution of the training data on different UEs. In the considered FL system, the model server faces a limited number of UEs to perform a FL task. If the server needs special datasets owned by certain UEs to train the learning model, the RS of different UEs is weak, namely $v$ tends to 0; otherwise, the learning model of the server can be trained on any dataset owned by the UEs, namely, the datasts owned by different UEs have a strong RS, so $v$ tends to 1.
    
    \item The issue of statistical heterogeneity of training data is not discussed here as it is beyond the scope of this paper. In the simulation, we set $v=0.5$ as in \cite{niyato} to represent a general case.  In \cite{niyato}, the authors presented the similar RS-based utility function to address the spectrum allocation problem in multi-user cognitive radio networks. Readers can refer to the first formula in Sec. IV.A, page 4275 of \cite{niyato} for detail.
    \end{enumerate}
\end{enumerate}

\section{Solution Of The Game}
We first solve problems (13) and (14), respectively, and obtain the NE of the game. Then, an iterative algorithm is developed to find
the NE in a distributed manner.

\subsubsection{Solve the MO problem \eqref{minU}} In the Bertrand game, each UE $k$ first gives the resource price $\Gamma_k$. Then, the MO can find the optimal resource purchase $\Theta$ according to $\Gamma_k,\ \forall k \in \mathcal{K}$, by solving problem \eqref{minU}. Through Taylor expansion, one can get $-\log\left(1-x\right) = x+\mathcal{O}\left(x\right)$. Then, Eq. \eqref{localiter} can be rewritten as
\begin{equation}
    I_k=-\eta_k\log(1-(1-\theta_k))= \eta_k\left(1-\theta_k\right)+\mathcal{O}(1-\theta_k). \label{changelocal}
\end{equation} 
By substituting eq. \eqref{changelocal} into eq. \eqref{minU}, $U_\text{S}$ can be rewritten as
\begin{alignat}{3}
  \tilde{U}_{\text{S}}= \sum_{t\in \mathcal{T}}\sum_{k\in K}\varrho_{k,t}\eta_k\left(1-\theta_k\right) +\frac{1}{2}\Big(\sum_{k\in K}(\theta_k)^2 +2v\sum_{k \neq j}(\theta_k \theta_j )\Big). \label{minX} 
\end{alignat}
Since $\tilde{U}_\text{S}(\theta_k)$ is convex with respect to $\theta_k$, one can differentiate $\tilde{U}_\text{S}(\theta_k)$ with respect to $\theta_k$ and set the results to 0. Then, the optimal solution of problem \eqref{minU} is obtained as
\begin{align}
  \textstyle \theta_k=A \sum_{t\in \mathcal{T}} \varrho_{k,t}+ B \sum_{j \in \mathcal{K},  j\neq k}\sum_{t\in \mathcal{T}}\varrho_{j,t}, \label{resource}
\end{align}
where $A=\frac{-(1-2v+Kv)}{(1-v)(Kv+1-v)}$ and $B=\frac{v}{(1-v)(Kv+1-v)}$ are constants.

\subsubsection{Solve the UE problem \eqref{maxUK}} After solving problem \eqref{minU}, the resource purchase $\Theta$ of the MO is known. Then, any UE $k$ can update its price $\Gamma_k$ by solving problem \eqref{maxUK}. Note that the pricing of UE $k$ for session $t$, i.e., $\varrho_{k,t}$ is affected not only by the accuracy $\theta_k$ but also by the pricing of the other UEs, which is represented by $\varrho_{-k,t}=\{\varrho_{x,t}|\forall x \in \mathcal{K}\ \text{and}\ x \neq k \}$. By substituting eq. \eqref{resource} into problem \eqref{maxUK} and expanding it with Taylor. The expression of $U_k$ can be rewritten as
\begin{align}
    & \textstyle \tilde{U}_k=\sum_{t\in\mathcal{T}} U_{k,t}, \label{uktaile}
    \end{align}
where $U_{k,t}=\varrho_{k,t} (1+A\sum_{t\in\mathcal{T}}\varrho_{k,t}-BV)-C_t(1+A\sum_{t\in\mathcal{T}} \varrho_{k,t}-BV)-{(D(1+A\sum_{t\in\mathcal{T}}\varrho_{k,t}-BV)^2+\tilde{E}_{k,t}^{\text{C}}})$.

As UE $k$ can estimate $f_{k,t}^\text{ex}$ and $g_{k,t}$ in each session $t$ by using eqs. \eqref{etl_state} and \eqref{chgain} respectively, the parameters $C_t=2\nu_kc_k|\mathcal{D}_k|f_{k,t}^\text{ex}$, $D=\frac{\nu_k(c_k)^2|\mathcal{D}_k|^2}{T^{\text{trn}}}$, $V=\sum_{x \in \mathcal{K}, x\neq k}\sum_{t\in\mathcal{T}}\varrho_{x,t}$ and $\tilde{E}_{k,t}^{\text{C}}$ in eq. (16) can be treated as constants by UE $k$.

Eq. \eqref{uktaile} shows that any UE $k$ can decompose the pricing problem into $I^\text{g}$ sub-problems, which can be solved in parallel to get $\varrho_{k,t}$ in each session $t$. Since $U_{k,t}(\varrho_{k,t})$ given in eq. \eqref{uktaile} is concave with respect to $\varrho_{k,t}$, one can differentiate $U_{k,t}(\varrho_{k,t})$ with respect to $\varrho_{k,t}$ and set the results to 0. Then, the solution of $\varrho_{k,t}$ is obtained as
\begin{align}
\textstyle \varrho_{k,t}=\frac{1-AC_{t}-BV-2AD+2ABDV}{2A^2D-2A}-\sum_{i \in \mathcal{T},  i \neq t} \varrho_{k,i}. \label{price}
\end{align}

\subsubsection{Find the NE of the game} At the NE of the game, neither the MO nor each of the UEs can get higher profit from changing the strategy profile, i.e., $\Theta$ or $\Gamma_k$.
Denote the NE strategy profile of the MO and the UEs by $\{\hat{\Theta},\hat{\Gamma}_1,\cdots,\hat{\Gamma}_k\}$, where $\hat{\Theta}=\left(\hat{\theta}_1,\cdots,\hat{\theta}_{\left|\mathcal{K}\right|}\right)$ and $\hat{\Gamma}_k=\left(\hat{\varrho}_{k,1},\cdots,\hat{\varrho}_{k,I^{g}}\right)$. We propose an iterative algorithm to find the NE. 

Let $i=1,2,\cdots$ denote the iteration numbers. In the $i^{\text{th}}$ iteration, we use $\Gamma_k[i]$, $\Gamma_{-k}[i]$ and $\Theta[i]$ to represent the pricing of UE $k$, the pricing of the UEs in set $\mathcal{K}$ other than UE $k$, and the resource purchase of the MO, respectively. Given that the pricing of all UEs are observable in a market, the following distributed algorithm is designed to find the NE.
\begin{algorithm}
\caption{Find the NE of the proposed Bertrand game.}

\begin{algorithmic}[1]
\STATE The MO specifies and broadcasts its performance parameters: $\epsilon$, $I^{\text{g}}$, $T^{\text{trn}}$ and $T^{\text{com}}$. 
\STATE Let $i=1$. Each UE $k$ initialize the price as $\Gamma_k[i]$.
\STATE The MO obtain the resource purchase $\Theta[i]$ using eq. \eqref{resource}.
\REPEAT
\STATE For each UE $k$, $\forall k \in \mathcal{K}$, after collecting $\theta[i]$ of the MO and $\Gamma_{-k}[i]$ of the other UEs in set $\mathcal{K}$, it can update the optimal pricing $\Gamma_k[i]$ by using eq. \eqref{price}.
\STATE For the MO, after collecting all $\Gamma_k[i]$, it can update the optimal resource purchase $\Theta[i]$ by using eq. \eqref{resource}.
\STATE Each UE $k$ derives the gradient $\nabla U_k(\varrho_{k,t})[i]$. 
\STATE Update $i=i+1$.
\UNTIL $\left\|\nabla U_k(\varrho_{k,t})[i]\right\| \leq \Xi \left\|\nabla U_k(\varrho_{k,t})[i-1]\right\|, \ \forall k\in{\mathcal{K}} $.

\end{algorithmic}

\end{algorithm}

Because $U_k(\varrho_{k,t})$ is strictly concave with respect to $\varrho_{k,t}$,
the upper bound on the iteration number of \textbf{Algorithm 1} to reach the convergence threshold $\Xi$ is $\mathcal{O}(\log\frac{1}{\Xi})$ \cite{2017Distributed} and the computational complexity of \textbf{Algorithm 1 } is $\log\frac{1}{\Xi}(K+I^g)$ \cite{Floating}. Although it is assumed that all the UEs are willing to engage in the FL task, some of them cannot be selected because their computing power fails to meet the requirement of local relative accuracy $\theta_k$. Then we apply constraint \eqref{positivel} to the outcome of \textbf{Algorithm 1} and determine which UEs are qualified for the FL task. Finally, we give the execution process of the game in the following \textbf{Algorithm 2}, which is termed as the \textit{task-load-aware game-theoretic scheme} (TLA-GTS)
\begin{algorithm}
    \renewcommand{\algorithmicrequire}{\textbf{Input:}}
    \renewcommand{\algorithmicensure}{\textbf{Output:}}
    \caption{TLA-GTS}
    \begin{algorithmic}[1]
    \REQUIRE \STATE The set of the potential client UEs $\mathcal{K}$ for the MO.
	\REPEAT
		\STATE Perform \textbf{Algorithm 1} and obtain the solution $\{\hat{\Theta},\hat{\Gamma}_k\}$.
		 \IF{$\theta_k \leq 0$ \textbf{or} $\theta_k > \theta^\text{max}$ (i.e., constraint \eqref{positivel} is not satisfied)}
		    \STATE Remove the $k^{\text{th}}$ UE  from set $\mathcal{K}.$ 
        \ENDIF
		\UNTIL $0<\theta_k\leq \theta^\text{max}$\ for $\forall k \in \mathcal{K}$, or $\mathcal{K}=\varnothing.$
		\ENSURE The remaining UEs in set $\mathcal{K}$.
  \end{algorithmic}
\end{algorithm}

In \textbf{Algorithm 2}, we initialize $\mathcal{K}$ as all the potential client UEs for the MO. After performing \textbf{Algorithm 1}, one can make the following choices according to the obtained solution. If the solution satisfies constraint \eqref{positivel}, the algorithm terminates and the solution is taken as the NE of the game. Otherwise, the UE with the highest price is removed from set $\mathcal{K}$, and \textbf{Algorithm 1} is performed again until constraint \eqref{positivel} is satisfied or set $\mathcal{K}$ is empty. An empty UE set indicates that the performance set by the MO (e.g., $\epsilon$, $I^{\text{g}}$, $T^{\text{trn}}$ and $T^{\text{com}}$) cannot be achieved by the UEs. The MO needs to lower the performance metrics and restart the game. \textbf{Algorithm 2} is implemented by performing \textbf{Algorithm 1} for $K$ times. Therefore, its computational complexity is $\log\frac{1}{\Xi}(K^2+KI^g)$.

\section{Simulation Results}
One MO and four UEs are placed in wireless networks, e.g., WLAN. We set $f^\text{max}=2$ GHz\cite{pingfangxiang} and discretize $f_{k,t}^\text{ex}$ ($0\leq f_{k,t}^\text{ex}\leq f^\text{max}$) of each UE $k$ into 5 levels. By using the statistical method, each UE $k$ can obtain a different state transition probability matrix $Q_k^\text{F}$ as follows.

\begin{equation*}
    Q_1^\text{F}=\begin{bmatrix}
     0.3 & 0.15 & 0.25 & 0.1 & 0.2\\
        0.2 & 0.1 & 0.1 & 0.4 & 0.2\\
        0.1 & 0.4 & 0.1  & 0.1 & 0.2\\
        0.4 & 0.1 & 0.1 & 0.2  & 0.2\\
        0.1 & 0.3 & 0.1 & 0.3  & 0.2
    \end{bmatrix}
    \end{equation*}
\begin{equation*}
   Q_2^\text{F}= \begin{bmatrix}
        0.2 & 0.3 & 0.1 & 0.1  & 0.3\\
        0.3 & 0.1 &  0.4& 0.1 & 0.1 \\
        0.2 & 0.2  &0.3  & 0.1 & 0.2\\
        0.1  & 0.3 & 0.4 &0.1  & 0.1\\
        0.1  & 0.2 & 0.4 &0.2  & 0.1
    \end{bmatrix}
\end{equation*}

\begin{equation*}
  Q_3^\text{F}= \begin{bmatrix}
        0.2 &0.1 &0.35 &0.15  & 0.1\\
        0.1 &0.15 &0.3 &0.25 &0.2 \\
        0.1 &0.1 &0.1  &0.4 &0.3\\
        0.1 &0.1 &0.4  &0.1  &0.3\\
        0.1 &0.1 &0.4  &0.3  &0.1
    \end{bmatrix}
    \end{equation*}
    \begin{equation*}
   Q_4^\text{F}= \begin{bmatrix}
        0.2 & 0.1 & 0.3 &0.2  & 0.2\\
        0.2 & 0.15 &  0.3& 0.25 &0.1 \\
        0.3 & 0.1  &0.2  & 0.1 & 0.3\\
        0.2 & 0.4 & 0.2 &0.1  & 0.1\\
        0.2 & 0.4 & 0.1 &0.2  &0.1
    \end{bmatrix} 
\end{equation*}

The lower and upper bounds of the time-varying channel gain $g_{k,\tau}$ is set to $\check{g}=2^{0.4}-1$ and $\hat{g}=2^{3.1}-1$\cite{2020Adaptive}, respectively, and the transition probability matrix $Q^\text{C}$ of $g_{k,\tau}$ as follows.
\begin{figure*}[ht]
\begin{equation*}
Q^{\text {C}}=\left(\begin{array}{llllllllll}
0.489 & 0.256 & 0.128 & 0.064 & 0.032 & 0.016 & 0.008 & 0.004 & 0.002 & 0.001 \\
0.001 & 0.489 & 0.256 & 0.128 & 0.064 & 0.032 & 0.016 & 0.008 & 0.004 & 0.002 \\
0.002 & 0.001 & 0.489 & 0.256 & 0.128 & 0.064 & 0.032 & 0.016 & 0.008 & 0.004 \\
0.004 & 0.002 & 0.001 & 0.489 & 0.256 & 0.128 & 0.064 & 0.032 & 0.016 & 0.008 \\
0.008 & 0.004 & 0.002 & 0.001 & 0.489 & 0.256 & 0.128 & 0.064 & 0.032 & 0.016 \\
0.016 & 0.008 & 0.004 & 0.002 & 0.001 & 0.489 & 0.256 & 0.128 & 0.064 & 0.032 \\
0.032 & 0.016 & 0.008 & 0.004 & 0.002 & 0.001 & 0.489 & 0.256 & 0.128 & 0.064 \\
0.064 & 0.032 & 0.016 & 0.008 & 0.004 & 0.002 & 0.001 & 0.489 & 0.256 & 0.128 \\
0.128 & 0.064 & 0.032 & 0.016 & 0.008 & 0.004 & 0.002 & 0.001 & 0.489 & 0.256 \\
0.256 & 0.128 & 0.064 & 0.032 & 0.016 & 0.008 & 0.004 & 0.002 & 0.001 & 0.489
\end{array}\right) .
\end{equation*}
\hrulefill  
\end{figure*}

The other simulation parameters include dataset size $|\mathcal{D}_k|=8\times10^7$ \cite{pingfangxiang}, the number of CPU cycles for computing one data sample $c_k= 15$\cite{pingfangxiang}, the size of model parameters $L=0.1$ Mb, the global iteration number $I^{\text{g}}=10$, the duration for each local training $T^{\text{trn}}=2$s, the duration for each parameter transmission $T^\text{com}=0.2$s, the CPU parameter $\nu_k=10^{-28}$\cite{pingfangxiang}, the bandwidth allocated to each UE $W=1$ MHz, the noise power $\sigma^2=10^{-9}$, the BER requirement $10^{-3}$, the substitutability factor for the MO $v=0.5$, $\eta_k=1$ in eq. (2), and $\zeta=1$ in eq. (3).

\begin{figure}[H]
  \centering
  \includegraphics[scale=0.5]{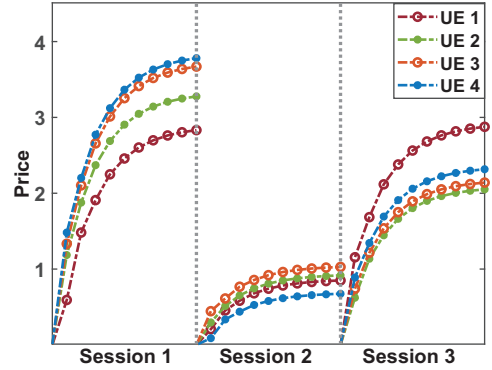}
  \caption{Price convergence.}
  \label{different_time_price}
\end{figure}

First, we show the impact of different EX-task load $f_{k,t}^\text{ex}$ on the pricing of the UEs. For that purpose, we set $f_{k,t=0}^\text{ex}=0$ for all the 4 UEs. Following eq. \eqref{etl_state}, $f_{k,t}^\text{ex}$ of the UEs in 3 consecutive training sessions (e.g., $t=1,2,3$) are estimated as $f_{1,t}^\text{ex}=\{0,0.5,1.5\}$, $f_{2,t}^\text{ex}=\{0.5,1,1\}$, $f_{3,t}^\text{ex}=\{1,1.5,1\}$,
and $f_{4,t}^\text{ex}=\{1,0,1\}$, respectively. Fig. \ref{different_time_price} shows the convergence of the prices of the UEs in the 3 training sessions.

As shown in Fig. \ref{different_time_price}, the UEs with lower $f_{k,t}^\text{ex}$ have a price advantage over the UEs with higher $f_{k,t}^\text{ex}$ in each training session. As the energy cost for data uploading $E_{k,t}^{\text{C}}$ is negligible compared to that for model training $E_{k,t}^{\text{F}}$, the results imply that the pricing of a UE is sensitive to its available computing power. To incentive the UEs with higher $f_{k,t}^\text{ex}$ (i.e. lower-level of available computing power) to engage in an FL task, the MO has to pay a higher price for the resource usage.
\begin{figure}[ht]
  \centering
  \includegraphics[scale=0.5]{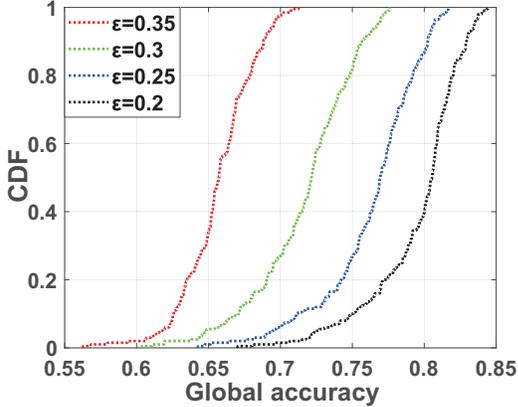}
  \caption{The global accuracy achieved by the TLA-GTS.}
  \label{cdf}
\end{figure}
Next, we show the performance of the proposed TLA-GTS, i.e.,
\textbf{Algorithm 2} on a real FL task. The task is to classify handwritten digits using the \texttt{MNIST} dataset \cite{726791}.
The distribution of the dataset over the UEs are balanced but non-i.i.d, and the training model is a 2-layer deep neural networks (DNN), consisting of one flatten layer, two fully-connected layers with rectified linear unit (ReLU) activation and a log-softmax output layer. For each ordered model accuracy \{.65, .7, .75, .8\}, 100 experiments (game cycles) were performed. Fig. \ref{cdf} shows the empirical cumulative distribution function (CDF) of the 100 experiments. One can see that the TLA-GTS can effectively motivate the autonomous UEs to help the MO achieve the accuracy goal of the FL task.

\begin{figure}[ht]
  \centering  \includegraphics[scale=0.7]{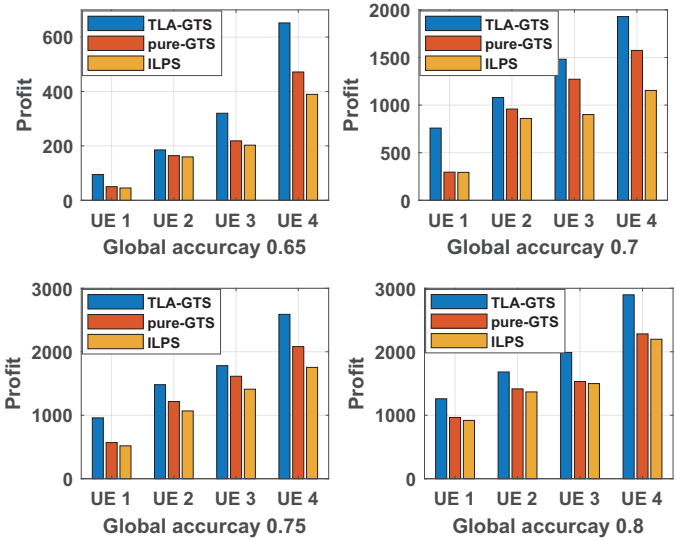}
  \caption{\small{The profits of the UEs using different pricing strategies.}} 
  \label{profitfig}
\end{figure}

Finally, we verify the superiority of TLA-GTS by comparing with other two schemes.
Given that most of the related works do not consider the user's EX-task load, the first scheme to be compared is the \textit{game-theoretic scheme without the task-load-aware mechanism}. It is termed as the \textit{pure}-GTS, which still uses the proposed game framework, but the energy cost of a UE does not consider the impact of its EX-task load, i.e., $\tilde{E}_{k,t}^{\text{F}}=\nu_k \left(f_k\right)^2T^{\text{trn}}$ in eq. (\ref{energytrn_2}). Another scheme to be compared is the \textit{independent linear pricing strategy} (ILPS), in which, each UE prices the MO independently using a linear pricing strategy based on its available resources, regardless of the pricing of other UEs. For each ordered model accuracy, the profits of the UEs using the different schemes are shown in Fig. \ref{profitfig}. From Fig. \ref{profitfig}, one can see that both the game-theoretic schemes TLA-GTS and \textit{pure}-GTS outperform the ILPS as they bring higher profits for each of the UEs. This is because  the game-theoretic schemes can price the resources of UEs based on their energy cost and the pricing of other UEs in the same training session. Furthermore, one can also note that the TLA-GTS outperforms the \textit{pure}-GTS in all cases. This is because that the energy cost of a UE in an FL task depends not only on the FL-task load but also on the EX-task load, which causes the UEs to produce the same amount of CPU cycles but consume different amount of energy. Hence, the TLA-GTS brings the UEs the highest profits in the game.

\section{Conclusion}
This paper has proposed a game theoretic approach to motivate UEs to participate in FL and the NE of the game has been obtained by a distributed iterative algorithm. Simulation results have verified the effectiveness of the proposed approaches.

\normalem
\bibliographystyle{IEEEtran}
\bibliography{IEEEabrv}

\begin{thebibliography}{10}
\providecommand{\url}[1]{#1}
\csname url@samestyle\endcsname
\providecommand{\newblock}{\relax}
\providecommand{\bibinfo}[2]{#2}
\providecommand{\BIBentrySTDinterwordspacing}{\spaceskip=0pt\relax}
\providecommand{\BIBentryALTinterwordstretchfactor}{4}
\providecommand{\BIBentryALTinterwordspacing}{\spaceskip=\fontdimen2\font plus
\BIBentryALTinterwordstretchfactor\fontdimen3\font minus
  \fontdimen4\font\relax}
\providecommand{\BIBforeignlanguage}[2]{{%
\expandafter\ifx\csname l@#1\endcsname\relax
\typeout{** WARNING: IEEEtran.bst: No hyphenation pattern has been}%
\typeout{** loaded for the language `#1'. Using the pattern for}%
\typeout{** the default language instead.}%
\else
\language=\csname l@#1\endcsname
\fi
#2}}
\providecommand{\BIBdecl}{\relax}
\BIBdecl

\bibitem{2016Federated}
{J. Konen et al}, ``Federated optimization: Distributed machine learning for
  on-device intelligence,'' 2016.

\bibitem{9409833}
Y.~Zhan, P.~Li, S.~Guo, and Z.~Qu, ``Incentive mechanism design for federated
  learning: Challenges and opportunities,'' \emph{IEEE Netw.}, vol.~35, no.~4,
  pp. 310--317, 2021.

\bibitem{8867906}
Y.~Sarikaya and O.~Ercetin, ``Motivating workers in federated learning: A
  stackelberg game perspective,'' \emph{IEEE Networking Letters}, vol.~2,
  no.~1, pp. 23--27, 2020.

\bibitem{8963610}
{Y Zhan et al}, ``A learning-based incentive mechanism for federated
  learning,'' \emph{IEEE Internet Things J.}, vol.~7, no.~7, pp. 6360--6368,
  2020.

\bibitem{8995775}
{S. R. Pandey et al}, ``A crowdsourcing framework for on-device federated
  learning,'' \emph{IEEE Trans. Wirel. Commun.}, vol.~19, no.~5, pp.
  3241--3256, 2020.

\bibitem{1988game}
D.~W. Majerus, ``Price vs. quantity competition in oligopoly supergames,''
  \emph{Econ. Lett.}, vol.~27, no.~3, pp. 293--297, 1988.

\bibitem{decentralized}
{W. Y. B. Lim et al}, ``Decentralized edge intelligence: A dynamic resource
  allocation framework for hierarchical federated learning,'' \emph{IEEE Trans.
  Parallel Distrib. Syst.}, vol.~33, no.~3, pp. 536--550, 2022.

\bibitem{dynamicgame}
W.~Y.~B. Lim, J.~S. Ng, Z.~Xiong, D.~Niyato, C.~Miao, and D.~I. Kim, ``Dynamic
  edge association and resource allocation in self-organizing hierarchical
  federated learning networks,'' \emph{IEEE J. Sel. Areas Commun.}, vol.~39,
  no.~12, pp. 3640--3653, 2021.

\bibitem{fsmccom}
S.~Salamat, B.~Khaleghi, M.~Imani, and T.~Rosing, ``Workload-aware
  opportunistic energy efficiency in multi-fpga platforms,'' in \emph{2019
  ICCAD}, 2019, pp. 1--8.

\bibitem{fsmccommunication}
{J. G. Ruiz et al}, ``On finite state markov chains for rayleigh channel
  modeling,'' in \emph{2009 Wireless VITAE}, 2009, pp. 191--195.

\bibitem{pingfangxiang}
T.~H. Thi~Le, N.~H. Tran, Y.~K. Tun, M.~N.~H. Nguyen, S.~R. Pandey, Z.~Han, and
  C.~S. Hong, ``An incentive mechanism for federated learning in wireless
  cellular networks: An auction approach,'' \emph{IEEE Trans. Wirel. Commun.},
  vol.~20, no.~8, pp. 4874--4887, 2021.

\bibitem{2021resource}
{W. Shu et al}, ``Resource demand prediction of cloud workloads using an
  attention-based gru model,'' in \emph{2021 17th International Conference on
  Mobility, Sensing and Networking (MSN)}, 2021, pp. 428--437.

\bibitem{2013tongji}
J.~Fenech, Y.~Yap, and S.~Shafik, ``\BIBforeignlanguage{English}{Brief
  technical note: A markov chain approach to measure investment rating
  migrations},'' \emph{\BIBforeignlanguage{English}{Australasian Accounting
  Business and Finance Journal}}, vol.~7, no.~3, pp. 145 -- 154, 2013.

\bibitem{tidaicanshu}
\BIBentryALTinterwordspacing
N.~Singh and X.~Vives, ``Price and quantity competition in a differentiated
  duopoly,'' \emph{RAND Journal of Economics}, vol.~15, no.~4, pp. 546--554,
  1984. [Online]. Available:
  \url{https://EconPapers.repec.org/RePEc:rje:randje:v:15:y:1984:i:winter:p:546-554}
\BIBentrySTDinterwordspacing

\bibitem{substitutability}
W.~S. Oh and R.~Muneepeerakul, ``How do substitutability and effort asymmetry
  change resource management in coupled natural-human systems?'' \emph{Palgrave
  Communications}, vol.~5, no.~1, pp. 1--8, 2019.

\bibitem{niyato}
D.~Niyato and E.~Hossain, ``Market-equilibrium, competitive, and cooperative
  pricing for spectrum sharing in cognitive radio networks: Analysis and
  comparison,'' \emph{IEEE Transactions on Wireless Communications}, vol.~7,
  no.~11, pp. 4273--4283, 2008.

\bibitem{2017Distributed}
{C. Ma et al}, ``Distributed optimization with arbitrary local solvers,''
  \emph{Optim. Method Softw.}, 2017.

\bibitem{Floating}
R.~Hunger, ``Floating point operations in matrix-vector calculus.''

\bibitem{2020Adaptive}
{ Y. Guo et al}, ``Adaptive bitrate streaming in wireless networks with
  transcoding at network edge using deep reinforcement learning,'' \emph{IEEE
  Trans. Veh. Technol.}, vol.~PP, no.~99, pp. 1--1, 2020.

\bibitem{726791}
{Y. Lecun et al}, ``Gradient-based learning applied to document recognition,''
  \emph{Proc. IEEE}, vol.~86, no.~11, pp. 2278--2324, 1998.

\end{thebibliography}

\end{document}